# Characterization of Request Sequences for List Accessing Problem and New Theoretical Results for MTF Algorithm

Rakesh Mohanty
Dept. of Comp. Sc. & Engg.
Indian Institute of Technology
Madras, Chennai, India

Burle Sharma
Dept. of Comp. Sc. & Engg
Sambalpur University Institute
of Information Technology
JyotiVihar, Burla, Orissa, India

Sasmita Tripathy
Dept. of Comp. Sc. & Engg
Veer Surendra Sai Institute of
Technology
Burla, Orissa, India

## ABSTRACT
List Accessing Problem is a well studied research problem in the context of linear search. Input to the list accessing problem is an unsorted linear list of distinct elements along with a sequence of requests, where each request is an access operation on an element of the list. A list accessing algorithm reorganizes the list while processing a request sequence on the list in order to minimize the access cost. Move-To-Front algorithm has been proved to be the best performing list accessing online algorithm till date in the literature. Characterization of the input request sequences corresponding to practical real life situations is a big challenge for the list accessing problem. As far as our knowledge is concerned, no characterization for the request sequences has been done in the literature till date for the list accessing problem. In this paper, we have characterized the request sequences for the list accessing problem based on several factors such as size of the list, size of the request sequence, ordering of elements and frequency of occurrence of elements in the request sequence. We have made a comprehensive study of MTF list accessing algorithm and obtained new theoretical results for our characterized special class of request sequences. Our characterization will open up a new direction of research for empirical analysis of list accessing algorithms for real life inputs.

## 1. INTRODUCTION
In Computer Science, linear search is one of the simplest search algorithm to find a particular element in the linear list. In linear search, we search an element sequentially one by one in a fixed size unsorted linear list from the start of the list and move towards the end of the list till the requested element is found. The performance of this data structure can be enhanced by making it self organizing. Each time after accessing the requested element, we reorganize the list by performing exchanges of adjacent elements so that the frequently requested elements are moved closer to the front of the list, thereby reducing the access cost of subsequent elements. The whole problem of efficiently reorganizing and accessing the elements of the list for obtaining optimal cost is called as List Accessing Problem. An algorithm that accesses the sequence of elements in the list based on the current and past requests is called List Accessing Algorithm. A list accessing algorithm uses a cost model to define the way in which the cost is assigned to a requested element when it is accessed in the linear unsorted list.

### 1.1 Problem Statement
In a list accessing problem, we are given a list $L$ of $l$ distinct elements, and a request sequence $R$ of $n$ elements. $R = x_1, x_2, \ldots, x_n$ such that $x_i \in L$, and $i = 1, 2, \ldots, n$ and $n \geq l$. Each time we access the element $x_i$ from $R$ in list $L$, we incur some access cost. After each access, list $L$ is reorganized in order to process $R$ efficiently. When we rearrange the list, we incur some reorganization cost. The total cost for accessing an element in the list is the sum of the access cost and the reorganization cost. Our objective to minimize the total cost while processing a request sequence on the list.

### 1.2 Applications
The list accessing techniques have been extensively used for storing and maintaining small dictionaries. There are various applications in which a linear list is the implementation of choice. It is used for organizing the list of identifiers maintained by a compiler and for resolving collisions in a hash table. Another important application of list accessing techniques is data compression. Other uses of List Accessing Algorithms are computing point maxima and convex hulls in computational geometry. The List Accessing Problem is also significant in the application of self organizing data structures.

### 1.3 Related Work
The list accessing problem is of significant theoretical and practical interest for the last four decades. As per our knowledge, study of list accessing techniques was initiated by the pioneering work of McCabe[1] in 1965. He investigated the problem of maintaining a sequential file and developed two algorithms Move-To-Front(MTF) and Transpose. From 1965 to 1985, the list update problem was studied by many researchers [2], [3], [4], [5] under the assumption that a request sequence is generated by a probability distribution. Hester and Hirschberg[6] have provided an extensive survey of average case analysis of list update algorithms. The seminal paper by Sleator and Tarjan [7] in 1985 made the competitive analysis of online algorithms very popular. The first use of randomization and the demonstration of its advantage in the competitive analysis context was done by Borodin, Linial and Saks [8] with respect to metrical task systems in 1985. Bachrach et. al. have provided an extensive theoretical and experimental study of online list accessing algorithms in 2002 [9]. Angelpolous and et. al.[10] have shown that MTF outperforms all other list accessing algorithms for request sequence with locality of reference property.

## 1.4 Our Contribution
In this paper, we have characterized the request sequences for the list accessing problem based on several factors such as size of the list, size of the request sequence, ordering of elements and frequency of occurrence of elements in the request sequence. Our characterization and classification of request sequence is a novel method which will facilitate generation of different request sequence for modeling the real world inputs for the list accessing problem. Here we have made a comprehensive study of MTF list accessing algorithm and obtained new theoretical results for our characterized special class of request sequences.

## 1.5 Organization of Paper
The paper is organized as follows. Section II contains a description of cost models and list accessing algorithms as well as illustration of MTF algorithm. Section III contains characterization of request sequence based on list size, request sequence size, ordering of elements and frequency of occurrence of elements. Section IV contains the analytical results of MTF algorithm. Section V provides the concluding remarks and focus on the future research issues.

## 2. PRELIMINARIES
## 2.1 List Accessing Cost Models
A cost model basically defines the way in which the cost is assigned to an element when it is accessed in the linear unsorted list. The two most widely used cost models for the list accessing problem are Full Cost Model by Sleator and Tarjan and Partial Cost Model by Ambuhl. For the Standard full cost model, the cost for accessing a requested element is equal to the position of that element in the input list i.e. for accessing the $i^{th}$ element in the list, access cost is $i$. Immediately after an access, the accessed element can be moved any distance forward in the list without paying any cost. These exchanges cost nothing and are called *free exchanges*. For any exchange between two adjacent elements in the list, cost is $1$. These exchanges are called *paid exchanges*. Hence total cost in a full cost model is the sum of number of paid exchanges and the access cost. For the partial cost model, the access cost is calculated by the number of comparisons between the accessed element and the elements present before the accessed element in the list. For accessing the $i^{th}$ element of the list, we have to make $i-1$ comparisons. Hence the access cost in partial cost model is $i-1$. The reorganization cost is same as the full cost model.

## 2.2 List Accessing Algorithms
There are two types of list accessing algorithms - online and offline. In online algorithms, the request sequence is partially known, i.e. we know the current request only and future requests come on the fly. In offline algorithms, we know the whole request sequence in advance. Till date many list accessing algorithms have been developed out of which the primitive algorithms are MTF, TRANSPOSE, and FC. In MTF, after accessing an element, the element is moved to the front of the list, without changing the relative order of the other elements. In TRANSPOSE, after accessing an element of the request sequence, it is exchanged with the immediately preceding element of the list. In FREQUENCY COUNT, we maintain a frequency count for each element of the list, each initialized to zero. We increase the count of an element by one whenever it is accessed. We maintain the list so that the elements are in non-increasing order of frequency count. It is proved that MTF algorithm is unique optimal algorithm for the list accessing problem. In our study, we have considered the Move-To-Front algorithm for the list accessing problem.

## 2.3 MTF Algorithm and Illustration
According to MTF algorithm "After accessing an element in the input list, it is move to front of the list, without changing the relative order of the other elements." We illustrate the MTF algorithm with the help of an example as follows. Let the list configuration is A B C D and request sequence is C A A D B. Each time after accessing a requested element in the list, the accessed element is moved to the front of the list, thereby shifting each of the preceding elements one position forward in the list. (This is shown in Table-1). Here, the total access cost for above input list and request sequence using MTF algorithm is 3+2+1+4+4=14.

**Table 1. Illustration of MTF algorithm**

| Steps | Accessed Element | List Configuration | Accessed Cost |
|---|---|---|---|
| 1 | C | A B (C) D | 3 |
| 2 | A | C (A) B D | 2 |
| 3 | A | (A) C B D | 1 |
| 4 | D | A C B (D) | 4 |
| 5 | B | D A C (B) | 4 |
| 6 | — | B D A C | 0 |
| | | | Total=14 |

## 3. CHARACTERISATION OF REQUEST SEQUENCES
For characterization of request sequences we have considered the following parameters.

(i) Size of the list – $l$

(ii) Size of the request sequence – $n$

(iii) A permutation representing the order of the elements in the list

(iv) Frequency of occurrence of element in the list.

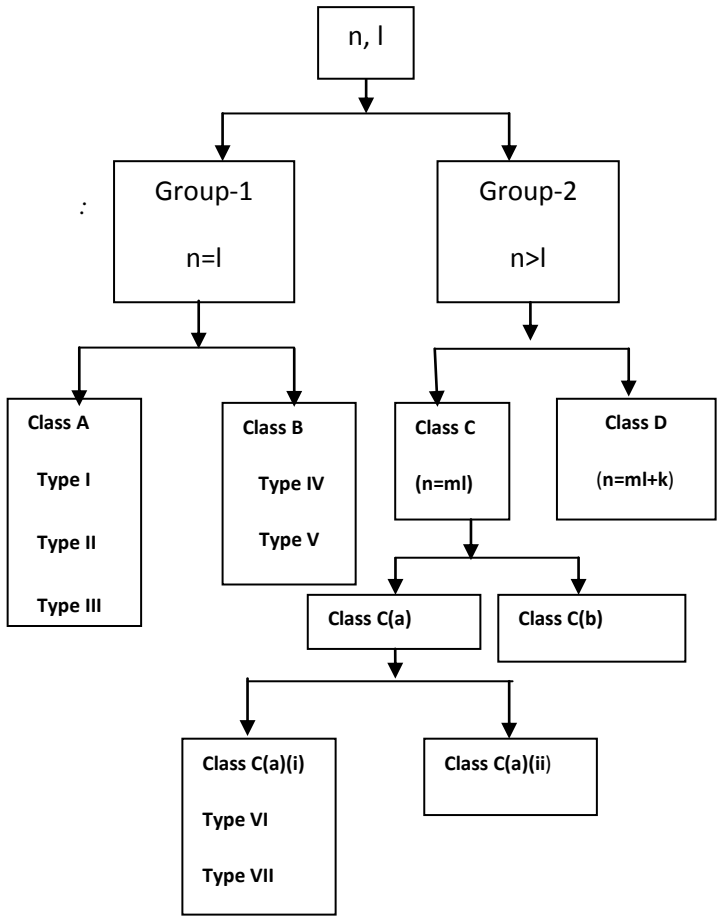

**Fig 1: Classification of Request Sequences**

We have classified the request sequences as shown in Fig. 1. Our characterization of request sequences is based on size of the request sequence and ordering of elements in the list with reference to the request sequence. Based on the comparison of size of the request sequence with size of the list, we can classify the request sequence into two groups. In Group 1, we consider the size of the request sequence is same as the size of the list. In Group 2, we consider the size of the request sequence is greater than the size of the list.

## 3.1. Characterization of Group 1

When the size of the request sequence is same as the size of the list (n=l), we classify the request sequence based on occurrence of elements in the list and request sequence into two different types – *Class A* and *Class B*. In Class A, all the elements of list must be present in the request sequence. Class A request sequence can be characterized as follows.

*Type I :* Request sequence is exactly the same as that of the list.
*Type II :* Request sequence is the reverse order as that of the list.

*Type III :* Request sequence is a permutation of arbitrary order as that of the list (except Type I and Type II).

In Class B, all elements of the list may not be present in the request sequence. Class B request sequence can be characterized as follows.

*Type IV :* Request sequence consist of any single element of the list at position p repeated n times where $1 \leq p \leq n$.

*Type V :* Request sequence consist of more than one elements each repeated at least once.

## 3.2. Characterization of Group 2

When the size of the request sequence is greater than the size of the list ($n > l$), Again we classify the request sequence based on the size of the request sequence along with size of the list into two different types - *Class C* and *Class D*. In Class C, size of the request sequence is a multiple of the size of the list. Class C request sequence can be characterized as follows. Class C(a) : all elements of the list must be present in the request sequence. Class C(b) : all elements of the list may not be present in the request sequence. Class C(a) request sequence can be characterized based on the frequency of elements occurrence in the request sequence as follows –

Class C(a)(i) : Frequency of all elements in the request sequence must be same.

*Type VI :* Type I data appear m number of times

*Type VII* : Type II data appear m number of times

Type C(a)(ii) : Frequency of elements in the request sequence may not be same

In Class D, size of the request sequence is not a multiple of the size of the list

## 4. RESULTS FOR MTF ALGORITHM
### 4.1 Assumptions
Let $l$ be the size of the list and $n$ be the size of the request sequence. The elements of the request sequence are considered to be distinct. We consider $n = l$ and Full Cost Model and Singly Linked List for our analysis of MTF.

*Illustration :* Let the List be 1,2,3. A request sequence with repetition of $2^{nd}$ elements 4 times will be 2,2,2,2. So, let $k = 2$ and $n = 4$. Then the cost for the above sequence when processed using MTF algorithm is 5 i.e. $n + k - 1$ = 4+2-1=5

## 4.2. Theoretical Results

*Theorem 1: MTF always gives best performance for a request sequence of size n of distinct elements, where the order of elements of the request sequence is same as that of the order of the list. The best case cost of MTF algorithm using FCM is denoted by $C_{MTF}(best) = \frac{n(n+1)}{2}$.*

Proof : Let $l$ be a list with $n$ elements $l_1$, $l_2$,….. $l_n$. Let $r$ be a request sequence with elements $r_1$, $r_2$,….. $r_n$ such that $r_1 = l_n$, $r_2 = l_{n-1}$……, $r_n = l_1$. Let $P_n$ be the best case cost for serving $r$ on $l$ using MTF. $P_n = \frac{n \times (n+1)}{2}$. We will prove this using induction.

Base : $P_1 = \frac{1 \times (1+1)}{2} = 1$. Let there is a single element in the list $l$ i.e. $l_1$ and single element in the request sequence $r$ i.e. $r_1$ when $r_1$ is served on $l_1$ using MTF the access cost is 1. Hence, $P_1$ is true.

Induction step : Let $P_n$ be true for $n = k$ i.e. best case cost $P_k = \frac{k \times (k+1)}{2}$. Now we have to prove by induction that $P_{k+1} = \frac{(k+1) \times ((k+1))}{2}$.

Let the elements of the list of size $k$ be $l_1$, $l_2$,….. $l_k$ and the elements of request sequence be $r_1$, $r_2$,….. $r_k$ such that the all the elements of the list are present in the request sequence. Let the $(k+1)^{th}$ element $l_{k+1}$ occurs after $l_k$ in the list and $r_{k+1}$ occur after $r_k$ in the request sequence. The access cost of $k$ elements of the request sequence is $\frac{k \times (k+1)}{2}$ where as the access cost of $(k+1)^{th}$ element of request sequence in the list is $k+1$. Hence, the total cost of serving $k+1$ elements in the request sequence is $\frac{k \times (k+1)}{2} + (k+1) = \frac{k(k+1)+2(k+1)}{2} = \frac{(k+1)(k+2)}{2} = \frac{(k+1)((k+1)+1)}{2} = P_{k+1}$. Hence the statement is true for all $n$. □

**Illustration**

Let the List be 1,2,3. A request sequence of equal size and distinct elements will be one of the following permutations of the list - 123, 132, 213, 231, 312, 321. Cost for the above request sequence, when processed using MTF algorithm are 6, 7,7,8,8,9 respectively. So, the worst case cost is found to be 9 i.e. $3^2$. Similarly, by increasing the size of the list and request sequence we can observe that the worst case cost for MTF will be $n^2$ for a request sequence of size $n$.

*Theorem 2: MTF always gives worst performance for a request sequence of size n of distinct elements, where the order of elements of the request sequence is that of the reverse order of the list. The worst case cost of MTF algorithm using FCM is denoted by $C_{MTF}(worst) = n^2$.*

Proof : Let $l$ be a list with $n$ elements $l_1$, $l_2$,….. $l_n$. Let $r$ be a request sequence with elements $r_1$, $r_2$,….. $r_n$ such that $r_1 = l_n$, $r_2 = l_{n-1}$……, $r_n = l_1$. Let $P_n$ be the worst case cost for serving $r$ on $l$ using MTF. $P_n = n^2$. We will prove this using induction.

Base : $P_1 = 1^2 = 1$. Let there is a single element in the list $l$ i.e. $l_1$ and single element in the request sequence $r$ i.e. $r_1$ when $r_1$ is served on $l_1$ using MTF the access cost is 1. Hence, $P_1$ is true.

Induction step : Let $P_n$ be true for $n = k$ i.e. worst case cost $P_k = k^2$. Now we have to prove by induction that $P_{k+1} = (k+1)^2$.

Let the elements of the list of size $k$ be $l_1$, $l_2$,….. $l_k$ and the elements of request sequence be $r_1$, $r_2$,….. $r_k$ such that $r_1 = l_k$, $r_2 = l_{k-1}$, ……, $r_k = l_1$. Let the $(k+1)^{th}$ element $l_{k+1}$ occurs after $l_k$ in the list and $r_{k+1}$ occur before $r_k$ in the request sequence. When $r_1 = l_{k+1}$ is served, access cost of $r_1$ is $k+1$. Then according to MTF rule, $l_{k+1}$ is moved to the front of the list. Now, the list configuration becomes $l_{k+1}, l_1, l_2,...., l_k$. The remaining request sequence left to be served is $r_2 = l_k$, $r_3 = l_{k-1}$, …, $r_{k+1} = l_1$. After serving the first request sequence $l_{k+1}$ and moving it to the front of the list the access cost of subsequent $k$ elements in the list is increased by 1 each. Hence, the total cost of serving next $k$ elements i.e. from $r_2$ to $r_{k+1}$ in the list is $k^2 + k$. Therefore, the total cost of serving $k+1$ elements in the request sequence is $k+1+k^2+k = k^2 + 2k + 1 = (k+1)^2 = P_{k+1}$. □

*Corollary 1: Let $C_{MTF}(Type\text{-}III)$ denote the total access cost incurred by MTF algorithm for Type III request sequence. Then $\frac{n(n+1)}{2} < C_{MTF}(Type\ III) < n^2$.* □

**Illustration**

Let the List be 1,2,3. A request sequence with repetition of $2^{nd}$ elements 4 times will be 2,2,2,2. So, let $k = 2$ and $n = 4$. Then the cost for the above sequence when processed using MTF algorithm is 5 i.e. $n + k - 1 = 4 + 2 - 1 = 5$

*Theorem 3: For Type IV request sequence of size n, the cost of MTF is given by to $C_{MTF}(Type\ IV) = m + n - 1$.* □

Proof : Let $l$ be a list with $n$ elements $l_1$, $l_2$,….. $l_p$. Let $r$ be a request sequence with elements $r_1$, $r_2$,….. $r_k$ such that $r_1 = l_m, r_2 = l_m, ...., r_n = l_m$ where $m$ is having any position the list. Let $P_n$ be the access cost for serving $r$ on $l$ using MTF. The access cost will be $n + m - 1$ where $n$ will be the number of elements of the request sequence. This will be proved by using induction.

Base : $P_1 = 1 + 1 - 1 = 1$. Let there is a single element in the list $l$ i.e. $l_1$ and single element in the request sequence $r$ i.e. $r_1$

when $r_1$ is served on $l_1$ using MTF the access cost is 1. Hence, $P_1$ is true.

Induction step : Let $P_n$ be true for $n = k$ i.e. access cost $P_k = k + m - 1$. Now we have to prove by induction $P_{k+1} = (k+1) + m - 1$. The access cost of $k$ elements of request sequence for $m^{th}$ element of the list is $k + m - 1$. After accessing the $m^{th}$ element in the request sequence first time i.e. for $r_1$ the element is moved to the front of the list. So for next $r_2, r_3, \ldots, r_k$ access cost is 1 for each element of the request sequence. So the access cost for $r_{k+1}$ element will be 1. Hence the total cost for serving $k+1$ elements in the request sequence is $k + m - 1 + 1 = (k+1) + m - 1 = P_{k+1}$. Hence it is true for all $n$.

*Corollary 2 : For Type IV request sequence of size n, the best case cost for MTF algorithm is n and worst case cost is 2n − 1.*

Illustration

Let the List be 1,2,3. A request sequence of equal size and distinct elements will be one of the following permutations of the list - 123, 132, 213, 231, 312, 321. Cost for the above request sequence, when processed using MTF algorithm are 6, 7,7,8,8,9 respectively. Let for a request sequence 213 all the elements are repeated twice and forms a new request sequence as 221133. Then cost for this request sequence can be derived as 7+3(2-1)=10 where 7 is the cost original request sequence, 3 is the number of elements in the original request sequence and 2 be the number of times each element of the original request sequence is repeated.

*Theorem 4: Let the access cost of a request sequence having distinct elements is represented by C for a list with same number of elements as of request sequence. Then for a new request sequence of any order where each element of the request sequence is repeated k times, the total cost of MTF algorithm for processing the request sequence can be evaluated by using the following formula. $C_{MTF} = C + n(k − 1)$ where n is size of the list and k is the number of time each element of the request sequence is repeated.*

Proof : Let $l$ be a list with $n$ elements $l_1, l_2, \ldots l_n$. Let $r$ be a request sequence with elements $r_1, r_2, \ldots r_n$ such that the request sequence consist of all the elements of list. Let $C$ be the access cost for any request sequence $r_1, r_2, \ldots r_n$. Let each element of the request sequence be repeated for $k$ times then the access cost for the request sequence with repetition of elements $P_n$ be $C + n(k − 1)$ where $n$ is the size of original request sequence. We will prove this using induction.

Base : $C = 1$. Let there is a single element in the list $l$ i.e. $l_1$ and single element in the request sequence $r$ i.e. $r_1$ when $r_1$ is served on $l_1$ using MTF the access cost is 1. Hence, $P_1$ is true.

Let $P_n$ be true for $n = s$ i.e. worst case cost $P_s = C + s(k − 1)$. Now we have to prove by induction that $P_{s+1} = C + (s + 1)(k − 1)$ where $k$ be fixed. Let the elements of the list of size s be $l_1, l_2, \ldots l_s$ and the elements of request sequence be $r_1, r_2, \ldots r_s$ such that the request sequence consist all the elements of the list.. So for $n = s + 1$ with $k$ repetitions the request sequence will be $(r_1)^k, (r_2)^k, \ldots, (r_s)^k, (r_{s+1})^k$. So, the total access cost will be $P_s + (s + 1) + 1$ as upto request sequence $r_s$ the cost is $P_s$, the access cost of $s + 1$ element of the request sequence is $s + 1$, then for subsequent access the cost is 1.

The cost upto $r_s$ can be represented as $P_s = 1 + 1(k − 1) + 2 + 1(k − 1) + \cdots + s + 1(k − 1) = 1 + 2 + \cdots + s + s(k − 1)$
$= \frac{s(s+1)}{2} + s(k − 1)$----------------eqn(1)

Then the cost of $P_{s+1}$ can be represented as $P_{s+1} = P_s + (s + 1) + (k − 1) = \frac{s(s+1)}{2} + s(k − 1) + (s + 1) + (k − 1)$
$= \frac{s(s+1)+2(s+1)}{2} + (s + 1)(k − 1) = \frac{(s+1)(s+2)}{2} + (s + 1)(k − 1)$
$= \frac{(s+1)((s+1)+1)}{2} + (s + 1)(k − 1)$

Hence, the statement is true for $s + 1$. Now, we have to prove that the statement is true for all $k + 1$.

From eqn(1) the statement is true for $k$ i.e. $P_s = \frac{s(s+1)}{2} + s(k − 1)$. Hence for $k + 1$ repetition,

$P_s = 1 + 1((k + 1) − 1) + 2 + 1((k + 1) − 1 + \cdots + s + 1((K + 1) − 1 = 1 + 2 + \cdots + s + s((k + 1) − 1) = \frac{s(s+1)}{2} + s((k + 1) − 1)$

Hence, the statement is true for $k + 1$. So, the statement is true for all $n$ and $k$.

**Illustration**

Let the List be 1,2,3. A request sequence of equal size and distinct elements will be one of the following permutations of the list - 123, 132, 213, 231, 312, 321. Cost for the above request sequence, when processed using MTF algorithm are 6, 7,7,8,8,9 respectively. Let for a request sequence 213 each element is repeated 2,3 and 4 times respectively and forms a new request sequence as 221113333. Then cost for this request sequence can be derived as 7+(2-1)+(3-1)+(4-1)=13 where 7 is the cost original request sequence, 2,3 and 4 be the number of times each element of the original request sequence is repeated.

*Theorem 5: Let the access cost of a request sequence having distinct elements is represented by C for a list with same number of elements as that of request sequence. Then for a new request sequence of any order where each element of the request sequence is repeated $k_1, k_2, \ldots, k_n$ times respectively, then the total cost of MTF for processing the request sequence can be evaluated by using the following formula. $C_{MTF} = C + (k_1 − 1) + (k_2 − 1) + \cdots + (k_n − 1)$ where n is size of the list and k is the number of time each element of the request sequence is repeated.*

Proof : Let $l$ be a list with $n$ elements $l_1$, $l_2$, ….. $l_n$. Let $r$ be a request sequence with elements $r_1$, $r_2$, ….. $r_n$ such that the request sequence consist of all the elements of list. Let all the elements of the request sequence be repeated differently i.e. $r_1$, $r_2$, ….. $r_n$ be repeated for $k_1, k_2, ….. k_n$ times respectively. Then the access cost for the new request sequence with repetition of elements $P_n$ be $C + (k_1 - 1) + (k_2 - 1) + \cdots + (k_n - 1)$. We will prove this using induction.

Base : $C = 1$. Let there is a single element in the list $l$ i.e. $l_1$ and single element in the request sequence $r$ i.e. $r_1$ when $r_1$ is served on $l_1$ using MTF the access cost is 1. Hence, $P_1$ is true.

Let $P_n$ be true for $n = s$ i.e. worst case cost $P_s = C + s(k-1)$. Now we have to prove by induction that $P_{s+1} = C + (s+1)(k-1)$ where $k$ be fixed. Let the elements of the list of size $s$ be $l_1$, $l_2$, ….. $l_s$ and the elements of request sequence be $r_1$, $r_2$, ….. $r_s$ such that the request sequence consist all the elements of the list.. So for $n = s+1$ with $k_1, k_2, ….. k_n, k_{n+1}$ repetitions respectively the request sequence will be $(r_1)^{k_1}, (r_2)^{k_2}, ….., (r_s)^{k_n}, (r_s)^{k_{n+1}}$. So, the total access cost will be $P_s + (s+1) + 1$ as upto request sequence $r_s$ the cost is $P_s$, the access cost of $s+1$ element of the request sequence is $s+1$, then for subsequent access the cost is 1.

The cost upto $r_s$ can be represented as $P_s = 1 + 1 \times (k_1 - 1) + 2 + 1 \times (k_2 - 1) + \cdots + s + 1 \times (k_n - 1) = 1 + 2 + \cdots + s + (k_1 - 1) + (k_2 - 1) + \cdots + (k_n - 1) = \frac{s(s+1)}{2} + (k_1 - 1) + (k_2 - 1) + \cdots + (k_n - 1)$ ----------------eqn(1)

Let the statement is true for $k_1 + 1, k_2 + 1, …, k_n + 1$

Then the cost of $P_s$ can be represented as

$P_s = 1 + 1 \times ((k_1 + 1) - 1) + 2 + 1 \times ((k_2 + 1) - 1) + \cdots + s + 1 \times ((k_n + 1) - 1) = 1 + 2 + \cdots + s + ((k_1 + 1) - 1) + ((k_2 + 1) - 1) + \cdots + ((k_n + 1) - 1) = \frac{s(s+1)}{2} + ((k_1 + 1) - 1) + ((k_2 + 1) - 1) + \cdots + ((k_n + 1) - 1)$

Hence, the statement is true for all $s$ and $k_1 + 1, k_2 + 1, …, k_n + 1$ Now, we have to prove that the statement is true for all $s+1$ with $k_1, k_2, ….. k_n, k_{n+1}$ repetitions respectively

The access cost for $P_{s+1} = P_s + (s+1) + 1 \times ((k_n + 1) - 1)$

$= \frac{s(s+1)}{2} + (k_1 - 1) + (k_2 - 1) + \cdots + (k_n - 1) + (s+1) + ((k_n + 1) - 1) = \frac{s(s+1)+2(s+1)}{2} + (k_1 - 1) + (k_2 - 1) + \cdots + (k_n - 1) + +((k_n + 1) - 1) = \frac{(s+1)+((s+1)+1)}{2} + (k_1 - 1) + (k_2 - 1) + \cdots + (k_n - 1) + +((k_n + 1) - 1)$

So, the statement is true for all $n$ and $k$.

## 5. CONCLUDING REMARKS
Our characterization and classification of request sequence is a novel method which will facilitate generation of different request sequence for modeling real world input for the list accessing problem. Further characterization of request sequences can be done based on locality of reference and look ahead property of the input. This characterization can be used as an important tool for making comparative performance analysis of various list accessing algorithms. New improved list accessing algorithms can be designed in future for a specific class of request sequence. Each characterization corresponds to a specific real life application for the list accessing problem. New cost models can be developed based on characterization of request sequence. Based on our characterization, the best list accessing algorithm can be determined for different inputs. This characterization will help us in developing some new alternate performance matrix for list accessing algorithms. A new experimental set up can be designed which will cover a wide range of request sequence for measuring the performance of list accessing algorithms.

## 6. ACKNOWLEDGMENTS

Our special thanks to Dr. N. S. Narayanaswamy of Department of Computer Science and Engineering, Indian Institute of Technology, Madras for his initial motivation and support.